\documentclass{article}
\usepackage[utf8]{inputenc}
\usepackage{authblk}
\usepackage{amsmath}
\usepackage{amssymb}
\usepackage{verbatim}
\usepackage{amsthm,amsfonts,bm}
\usepackage{MnSymbol}
\usepackage{cite}
\usepackage{graphicx}
\usepackage{color}
\usepackage[colorlinks=true, allcolors=blue]{hyperref}

\definecolor{Red}{rgb}{0.9,0,0}
\newcommand{\p}{\partial}

\title{To square root the Lagrangian or not: \\ an underlying geometrical analysis on classical and relativistic mechanical models}

\author[1]{B. F. Rizzuti}
\author[1]{G. F. Vasconcelos Júnior}
\author[1]{M. A. Resende}
\affil[1]{Departamento de Física, Universidade Federal de Juiz de Fora, MG, Brazil}
\date{}                     
\begin{document}

\maketitle

\begin{abstract}
The geodesic has a fundamental role in physics and in mathematics: roughly speaking, it represents the curve that minimizes the arc length between two points on a manifold. We analyze a basic but misinterpreted difference between the Lagrangian that gives the arc length of a curve and the one that describes the motion of a free particle in curved space. They are taken as equivalent but it is not the case from the beginning. We explore this difference from a geometrical point of view, where we observe that the non-equivalence is nothing more than a matter of symmetry. The equivalence appears, however, after fixing a particular condition. As applications, some distinct models are studied. In particular, we explore the standard free relativistic particle, a couple of spinning particle models and also the forceless mechanics formulated by Hertz. 
\end{abstract}

\section{Introduction}

Differential geometry is extensively used as the basic tool for description of many distinct areas of physics, after all it is widely accepted the space-time possesses the very structure of a differentiable manifold \cite{sen}. Besides that, it goes from special and general relativity theories \cite{o2006elementary}, till electrodynamics and gauge theories \cite{nakahara}. We would like to focus our attention to its application even in the less advanced topic, though not less important, of classical mechanics. One may find an extensive literature on the subject, see \cite{alexeibook, alexei} and references therein. The central idea of these notes is to draw attention to an underlying fact, which is often misinterpreted: given a (semi-)riemannian manifold $(\mathcal{M}, g)$, the action functionals\footnote{Here $V$ is a tangent vector. }
\begin{eqnarray}
S_1 = \int d\tau \sqrt{g(V,V)}
\end{eqnarray}
and
\begin{eqnarray}
S_2 = \int d\tau \frac{1}{2} g(V,V)
\end{eqnarray}
are not necessarily equivalent. In the literature, this fact is usually ignored. However, it is a subtle result that cannot be taken for granted.
This problem will become clear along the text and it is intimately related to what is known as pseudo-classical mechanics \cite{alexeipla}, that is, models whose number of configuration variables is less than the number of physical degrees of freedom. It turns out that these models are useful to describe quantum phenomena even before quantization, see \cite{bep, abgp}, justifying, thus, our presentation. 


The paper is divided as follows. In Section \ref{sec2}, we will show the non-equivalence between $S_1$ and $S_2$ and under what conditions they become equivalent. Sections \ref{sec3} and \ref{sec4} are dedicated to examples on the lack of equivalence between actions $S_1$ and $S_2$. In particular, we discuss in the former the free relativistic particle, while the latter contains examples of classical models of spinning particles. These consist of an example of a semiclassical model whose quantization leads to the Pauli equation \cite{aap} and  another model for the relativistic rotator, where the particle is considered to resemble a moving clock with the pointer characterizing its spin \cite{kuzenko, staruszkiewicz}. Then, we devote the entire Section \ref{sec5} to explore the forceless mechanics of Hertz \cite{barut1989geometry, alexeibook}. H. Hertz has provided a particular description of Newtonian classical mechanics without forces. His clever idea was to insert one more nonphysical dimension into the game, together with the introduction of the Newtonian potential inside the metric. In this case, second Newton law appears as the geodesic equation of a free particle in a curved manifold. These applications were chosen because they represent very clearly the difference between systems described by $S_1$ and $S_2$.  
Finally, Section \ref{con} is left for conclusions. Throughout all the paper, Einstein notation of sum over repeated indexes is adopted.

\section{$L$ or $\sqrt{L}$?}
\label{sec2}

Let us consider a (semi-)riemannian $m$-dimensional manifold $(\mathcal{M}^m, g)$ and a chart $(\varphi, U)$. $U$ is an open set $U \subset \mathcal{M}^m$ and 
\begin{eqnarray}
\varphi: U &\rightarrow& A(\mbox{open})\subset \mathbb{R}^m \cr P & \mapsto & \varphi(P):= (x^1(P),..., x^m(P)).
\end{eqnarray}
We write collectively 
\begin{eqnarray}
\varphi^i(P) = x^i(P); \, i=1,...,m
\end{eqnarray}
and elements of $\mathcal{M}^m$ will be denoted by capital letters. 

Let $\mathfrak{X}(\mathcal{M}^m)$ be the set of all smooth vector fields on $\mathcal{M}^m$, that is, the set of mappings $V$ which assigns to each point $P$ on $\mathcal{M}^m$ a tangent vector $V_P$ in the tangent space $T_P(\mathcal{M}^m)$. In coordinates $\varphi$, $V = V^i \partial_i$. We also fix the notation for the (pseudo-)metric $g$ in the coordinates $\varphi$,
\begin{eqnarray}
L:= g(V,V) = g(V^i \partial_i, V^j \partial_j) = g_{ij}V^i V^j.
\end{eqnarray}
For the case, there is a unique connection 
\begin{eqnarray}
D: \mathfrak{X}(\mathcal{M}^m) \times \mathfrak{X}(\mathcal{M}^m) \rightarrow \mathfrak{X}(\mathcal{M}^m), 
\end{eqnarray}
known as the Levi-Civita connection, such that in $\varphi$, it reads
\begin{eqnarray}\label{Def.SymbCrist}
D_{\partial_i}(\partial_j) = \Gamma^k{}_{ij}\partial_k; \quad
\Gamma^i{}_{jk} = \frac{1}{2}g^{il}(\partial_j g_{kl} + \partial_k g_{jl} - \partial_l g_{jk}).
\end{eqnarray}
$g^{il}$ denotes the elements of $g^{-1}$. A particle shall be denoted by the corresponding curve it traces \cite{o2006elementary}
\begin{eqnarray}
\alpha : I \subset \mathbb{R} & \rightarrow & \mathcal{M}^m \cr
\tau & \mapsto & \alpha(\tau).
\end{eqnarray}
$I$ is an open interval on $\mathbb{R}$. In this sense, $x^i(\tau)$ corresponds to the representation of $\alpha$ in the coordinates $\varphi$,
\begin{eqnarray}
(\varphi^i \circ \alpha)(\tau):= x^i(\tau),
\end{eqnarray}
as well as $\dot{x}^i(\tau)$ are the components of the tangent vector $\frac{d}{d\tau}$ in the point $\alpha(\tau)$
\begin{eqnarray}
\frac{d}{d\tau}\Bigg |_{\alpha(\tau)}:= \dot{x}^i(\tau)\partial_i | _{\alpha(\tau)}. 
\end{eqnarray}
We have denoted $\dot{x}^i \equiv \frac{dx^i}{d\tau}$. Since the state of motion is intuitively described by the corresponding position together with its velocity, we are naturally led to the notion of a tangent bundle, 
\begin{eqnarray}
T(\mathcal{M}^m) = \bigcup_{P  \in \mathcal{M}^m} T_P (\mathcal{M}^m)
\end{eqnarray}
whose elements are parametrized by $(x^{i}, \dot{x}^{i})$.

There are two common ways to find the equations of motion of the particle on the manifold. Although both of them are based on the principle of least action, they are fundamentally different. The equations of motion the distinct actions only become equivalent after a particular condition. The main objective of this work is to clarify this point. Let us consider the two actions, already presented in the Introduction, which are linear functionals defined on the space \begin{eqnarray}
\mathfrak{F}(T(\mathcal{M}^m)):= \{f:T(\mathcal{M}^m) \rightarrow \mathbb{R}; \, f \, \mbox{is smooth}\, \}. 
\end{eqnarray}
We then take two different smooth functions to write 
\begin{eqnarray}
S_1 = \int^{\tau_2}_{\tau_1} d\tau \sqrt{L}\equiv \int^{\tau_2}_{\tau_1} d\tau \sqrt{g_{ij}(x) \dot{x}^i \dot{x}^j}
\end{eqnarray}
and
\begin{eqnarray}
S_2 = \int^{\tau_2}_{\tau_1} d\tau \frac{1}{2}L \equiv \int^{\tau_2}_{\tau_1} d\tau \frac{1}{2} g_{ij}(x) \dot{x}^i \dot{x}^j.
\end{eqnarray}
In a sense, $S_1$ has a geometrical meaning, as it describes the arc length of the curve $\alpha$ connecting the points $\alpha(\tau_1)$ and $\alpha(\tau_2)$. On the other hand, $S_2$ is analogous to the usual action in classical mechanics of a free particle, for we are integrating a kinetic energy term. 

Let us discuss separately both actions, explaining why one or the other shall be elected for further investigations. 

\subsection{$S_1$:  geometric point of view}


Adopting the standard notion that $S_1$ is related to the arc length of $\alpha$ connecting $\alpha(\tau_1)$ and $\alpha(\tau_2)$, the extremization of $S_1$ provides the following differential equation for the curve which has minimum length 
\begin{eqnarray}\label{geo}
\Big(\frac{\dot{x}^i}{\sqrt{\dot{x}g\dot{x}}}\Big)^. + \frac{1}{\sqrt{\dot{x}g\dot{x}}}\Gamma^i{}_{jk}\dot{x}^j \dot{x}^k=0,
\label{eq_geral_S1}
\end{eqnarray}
in which we have used the short notation $\dot x g \dot x:= g_{ij} \dot{x}^i \dot{x}^j$. The coefficients $\Gamma$ were given in Eq. \eqref{Def.SymbCrist}. The subsequent analysis of this equation relies on how the quantity $\dot{x}g\dot{x}$ relates with the parameter $\tau$. We shall do it by separating this analysis in two distinct cases. First, we consider the case of the quantity $\dot{x}g\dot{x}$ as a function of $\tau$. Then, we treat it as a constant with respect to $\tau$.

Let us consider $\dot{x}g\dot{x}$ as a function of the parameter $\tau$. Thus, the time derivative in the first term of Eq. \eqref{eq_geral_S1} implies the equation
\begin{equation}\label{project.geo}
\Pi^i{}_j\Big[\ddot{x}^j + \Gamma^j{}_{kl}\dot{x}^k \dot{x}^l=0\Big],
\end{equation}
 where $\Pi^i{}_j = \delta^i{}_j - \frac{\dot{x}^i g_{jk}\dot{x}^k}{\dot{x}g\dot{x}}$, or, by denoting $\Lambda^i{}_j = \frac{\dot{x}^i g_{jk}\dot{x}^k}{\dot{x}g\dot{x}}$,
 \begin{equation}
     \Pi^i{}_j = \delta^i{}_j - \Lambda^i{}_j.
 \end{equation}
It follows that the vector $\dot{x}$, whose components are given by $\dot{x}^j$, is in the null space of the operator associated with the matrix $\Pi$. Thus, the null space is not trivial and $\Pi$ is not invertible, implying that $\text{det}\; \Pi = 0$. Besides that, the Hessian matrix is proportional to the matrix $\Pi$. 
\begin{equation}
    \frac{\partial^2L}{\partial\dot{x}^i\partial\dot{x}^j} \sim \Pi^i{}_j.
\end{equation}
Therefore, we are dealing with a singular Lagrangian theory. It can be shown that this is intimately related with the reparametrization invariance of this functional, which we will cover below. Before we consider the case where $\sqrt{\dot{x}g\dot{x}}$ does not depend of $\tau$, we can also understand the ambiguity of the singular system in terms of the algebraic properties of the operators $\Pi$ and $\Lambda$ \cite{alexeibook}. We point out that they are projectors because of the following 
\begin{eqnarray}
\begin{array}{cc}
     \Pi^2 = \Pi, \,\, \Lambda^2 = \Lambda  \\
     \Pi \Lambda = 0 \\
     \Pi + \Lambda = \mathbb{I}.
\end{array}
\end{eqnarray}
So, given an arbitrary vector $v$, it is uniquely decomposed as 
\begin{equation}
    v = \Pi v + \Lambda v :=v_{\perp}+v_{\shortparallel}.
\end{equation}
If we fix the vector $\dot{x}^i$ that specifies the projector $\Pi$, then 
\begin{equation}
    v^i_{\shortparallel} = \frac{g(\dot x, v)}{g(\dot x, \dot x)} \dot x^i;  \,\,\, g(\dot x, v_{\perp})=0. 
\end{equation}
We observe that $v_{\shortparallel}$ is a projection of $v$ on the direction of $\dot x^i$ whilst $v_{\perp}$ belongs to the corresponding orthogonal complement. Considering the vector $v$ defined by 
\begin{equation}
    v^i:= \ddot{x}^i + \Gamma^i{}_{jk} \dot x^j \dot x^k, 
\end{equation}
our variational problem has led us to the equation of motion \eqref{project.geo}, which is just the projection $\Pi v = v_{\perp} = 0$. This restricts $v_{\perp}$, but $v_{\shortparallel}$ remains arbitrary. This is the ambiguity exposed before.  

Now, we work with $\sqrt{\dot{x}g\dot{x}}$ as if it were a constant with respect to $\tau$. Then, it follows that Eq. \eqref{eq_geral_S1} turns in to
\begin{equation}\label{geo1}
	\ddot{x}^i + \Gamma^i{}_{jk}\dot{x}^j \dot{x}^k = 0.
\end{equation}
This is the well known equation of a geodesic. The problem is the meaning of the evolution parameter. In fact, $S_1$ is invariant under reparametrizations,

\begin{eqnarray}\label{ri}
\tau \rightarrow \tau' = f(\tau)
\end{eqnarray}
for any smooth function $f:\mathbb{R} \rightarrow \mathbb{R}$. This picture is understood as a lack of physical meaning of $\tau$, which, in turn, implies no physical meaning for the solution $x^i(\tau)$ of Eq. (\ref{geo}). This fact is known in the literature as pseudo-classical mechanics: the number of configuration variables is less than the number of physical degrees of freedom \cite{alexeipla}. The origin of reparametrization invariance is clarified if we interpret $\sqrt{L}= \sqrt{g_{ij} \dot{x}^i \dot{x}^j}$ as Lagrangian and use the Dirac-Bergmann algorithm of hamiltonization to constrained systems \cite{alexeibook, Lusanna_2018}. Defining the conjugate momenta to the variables $x^i$
\begin{eqnarray}
\pi_i:= \frac{\p \sqrt{L}}{\p \dot{x}^i} = \frac{g_{ij} \dot{x}^j}{\sqrt{\dot x g \dot x}},
\end{eqnarray}
one immediately finds the constraint
\begin{eqnarray}
g^{ij}\pi_i \pi_j = 1.
\end{eqnarray}

Invoking the Dirac conjecture that first class constraints generate a local symmetry \cite{aadbfr1}, we can write a local infinitesimal transformation that leaves the Lagrangian $\sqrt{L}$ invariant (modulo a total derivative term)
\begin{eqnarray}\label{sym}
\delta x^i = \frac{1}{2}\varepsilon (\tau) \{ x^i, g^{ij}\pi_i \pi_j -1 \}|_{\pi(x, \dot x)}.
\end{eqnarray}
In the above formula, the factor $1/2$ is set by convenience, $\varepsilon$ is an arbitrary smooth function of $\tau$ and $\{,\}$ is the Poisson bracket defined by
\begin{eqnarray}
\{A,B\} := \frac{\p A}{\p x^i}\frac{\p B}{\p \pi_i} - \frac{\p A}{\p \pi_i}\frac{\p B}{\p x^i}. 
\end{eqnarray}
Thus, (\ref{sym}) reads
\begin{eqnarray}\label{sym1}
\delta x^i = \varepsilon \frac{\dot x^i}{\sqrt{\dot x g \dot x}}.
\end{eqnarray}
A direct calculation shows that 
\begin{eqnarray}
\delta \sqrt{L} = \frac{d}{d \tau} \left ( \sqrt{g_{ij}\left( \frac{\varepsilon \dot{x}^i}{\sqrt{\dot{x}g \dot{x}}} \right )\left( \frac{\varepsilon \dot{x}^j}{\sqrt{\dot{x}g \dot{x}}} \right ) } \right ), 
\end{eqnarray}
as stated. 

The situation here is analogous to electrodynamics \cite{aadbfr1}. In the former case, equations of motion are derived from a least action principle applied to the Lagrangian 
\begin{eqnarray}
\mathcal{L} = -\frac{1}{4}F_{\mu \nu} F^{\mu \nu}
\end{eqnarray}
where $F_{\mu \nu} = \p_\mu A_\nu - \p_\nu A_\mu$ and $\p_\mu : = \frac{\p}{\p x^{\mu}}$. Clearly, the model possesses the local symmetry
\begin{eqnarray}\label{gau}
\delta A_\mu = \p_\mu \psi,
\end{eqnarray}
where $\psi$ is an arbitrary function of space-time variables. Since $A_\mu$ changes under the transformation (\ref{gau}), it is not an observable of the theory. The physical sector is obtained by looking for combinations of the field $A_\mu$ and its derivative that remain unchanged by (\ref{gau}). For example, $\delta F_{\mu \nu} = 0$ and $\delta (\oint \Vec{A}\cdot \vec{dl})=0$. $F_{\mu \nu}$ are just components of the electromagnetic field whereas $\oint \vec{A} \cdot \vec{dl}$ manifests itself through the Aharonov-Bohm effect \cite{Aharonov_Bohm_1959}. 

In our case, the coordinates $x^i$ are not all observables because they are altered by the local symmetry the model presents, see (\ref{sym1}). It is, though, possible to find the physical sector of the model and we shall do it for the particular case of a free relativistic particle, see Section \ref{sec3}.

\subsection{$S_2$: kinetic point of view}\label{kineticpov}

Starting now with the action $S_2$, one finds the following equations of motion 
\begin{eqnarray}\label{geo2}
\ddot{x}^i + \Gamma^i{}_{jk} \dot{x}^j \dot{x}^k = 0,
\end{eqnarray}
also after a least action principle. For this case, we have no local symmetry and there is no arbitrariness as the one posed by \eqref{ri}. Thus, $\tau$ is interpreted as the evolution parameter. On the contrary of $S_1$, if one writes the conjugate momenta 
\begin{eqnarray}\label{pi}
\pi_i = \frac{\p }{\p \dot x^i}\left (\frac{1}{2}L \right ) = g_{ij} \dot x^j, 
\end{eqnarray}
then (\ref{pi}) may be interpreted as an algebraic equation to obtain the maximum number of velocities $\dot x^i$ in terms of $x^i$ and $\pi_j$. It turns out that we may find all of them once $g$ is invertible,
\begin{eqnarray}
\dot x^i = g^{ij}\pi_j
\end{eqnarray}
and no constraint, such as $g^{ij} \pi_i \pi_j = 1$, appears. In this case, all coordinates $x^i$ of a particle have deterministic evolution given by (\ref{geo2}). Naturally, the equations of motion for both actions $S_1$ and $S_2$ \eqref{geo} and \eqref{geo2} are different. However, once $\sqrt{\dot x g \dot x}$ is fixed, the equations of motion coincide, see \eqref{geo1} and \eqref{geo2}. Hence, $S_1$ turns out to be equivalent to $S_2$. It is possible to see the meaning of $\tau$ in $S_1$ as well. With $\sqrt{\dot x g \dot x}=1$, $S_1 = \int^{\tau_2}_{\tau_1} d\tau = \tau_2 - \tau_1$, that is the action is proportional to $\tau$, which in this case may be called the evolution parameter \cite{wthirring}. 

We will consider in the next sections some examples to show the relevance of one or the other formulation.  

\section{Free Relativistic Particle}
\label{sec3}

The free relativistic particle is the canonical example where the reparametrization invariance plays a central role \cite{Fulop_1999}. For the case, let us consider Minkowski space-time $\mathbb{M}$ as our manifold. It is isometric to the semi-riemannian manifold $(\mathbb{R}^4_1, \eta)$. The index $'1'$ means that $-\eta$ is positive defined in a subspace of dimension $4-1=3$. We chose its signature to be 
$\eta= \mbox{diag}(+1,-1,-1,-1)$, providing the known Einstein-Weyl space-time causal structure. Let us take coordinates $x^{\mu}$. As usual, Greek letters $\mu$, $\nu$, etc run the values $0$, $1$, $2$ and $3$. $x^0$ is related to the time coordinate while $x^i$, $i=1,2,3$ are the spatial ones.\footnote{Actually, $\mathbb{M}$ is a set of events. $\mathbb{R}^4_1$ is the linear space with $\mathbb{M}$ being the corresponding affine space.}  

We begin our analysis by writing the action
\begin{eqnarray}
S_a = \int d\tau \frac{1}{2}\eta_{\mu \nu}\dot x^{\mu} \dot x^{\nu}.
\end{eqnarray}
The principle of least action provides the deterministic evolution
\begin{eqnarray}
\ddot x^{\mu} = 0 \Rightarrow x^\mu(\tau) = V^\mu \tau + X^\mu.
\end{eqnarray}
$V^\mu$ and $X^\mu$ are constants. We have then a straight line, with no physical interpretation though. In fact, the speed of the particle has no superior bound. It would be the case, if one sets a restriction of the form $\eta_{\mu \nu}\dot x^{\mu} \dot x^\nu = 1$, to impose, in turn, a maximum value to the modulus of the physical three velocity. So, we write
\begin{eqnarray}\label{L.Free_Relativistic_Particle}
S_b = \int d\tau \left ( \frac{1}{2e}\eta_{\mu \nu}\dot x^{\mu} \dot x^\nu + \frac{e}{2} \right ).
\end{eqnarray}
In this case, $e$ is an auxiliary variable. Its equation of motion is given by
\begin{eqnarray}
-\frac{1}{2e^2}\eta_{\mu \nu}\dot x^{\mu} \dot x^\nu +\frac{1}{2}= 0 \Rightarrow e = \sqrt{\eta_{\mu \nu}\dot x^{\mu} \dot x^\nu}.
\end{eqnarray}
This last expression shows that $e(\tau)$ has no independent dynamics. It is completely described by the evolution of $x^\mu$. If we substitute $\sqrt{\eta_{\mu \nu}\dot x^{\mu} \dot x^\nu}$ back into the action $S_b$ we finally get 
\begin{eqnarray}
S_c = \int d\tau \sqrt{\eta_{\mu \nu}\dot x^{\mu} \dot x^\nu},\label{sc}
\end{eqnarray}
which has the same structure of $S_1$. As discussed before, $S_c$ has invariance under reparametrizations. In this case $x^\mu(\tau)$ has no physical interpretation. $S_c$ has, on the other hand, global invariance under Lorentz transformation
\begin{eqnarray}
x^\mu \rightarrow x'^\mu = \Lambda^\mu{}_\nu x^\nu; \,\, \Lambda \in SO(1,3).
\end{eqnarray}
Actually, this is one of the main reasons why we work with non-physical degrees of freedom: it is possible to carry a fully Lorentz covariant model along the way, knowing that not all of the variables are physical ones. 

Now, the principle of least action provides the following equations of motion from $S_c$,
\begin{eqnarray}
\frac{\eta_{\mu \nu} \dot x^\nu}{\sqrt{\dot x \eta \dot x}} = \pi_{\mu} = \mbox{const.}
\end{eqnarray}
together with the constraint $\eta^{\mu \nu}\pi_\mu \pi_\nu=1$. The physical sector can be obtained if we eliminate the $\tau$-dependence of the game. It may be done by declaring $x^0$ as the evolution parameter. Then, the deterministic evolution of spatial variables $x^i$ in terms of $x^0$ is given by 
\begin{eqnarray}
\frac{dx^i}{dx^0} = \frac{\dot x^i}{\dot x^0} = \frac{\pi^i}{\pi^0} = \frac{\pi^i}{\sqrt{1+ (\pi^i)^2}}.
\end{eqnarray}
The last equality comes from the constraint: $\pi^0 = \sqrt{1+ (\pi^i)^2}$. Thus, we have a particle with bounded three-velocity walking, as expected, in a straight line. 

All the structure presented above is commonly used in different contexts. On the next Section, we will discuss how it appears for description of classical spinning particles. 

\section{Spinning Particles}
\label{sec4}

In this Section we discuss two more examples where invariance under reparametrization is present. In the first case, we discuss a particular model whose canonical quantization leads to the Pauli equation. The second treats a relativistic rotator, known as the Staruszkiewicz model.

\subsection{Classical spinning particle}
Let us now see how this formalism is applied to the construction of a spinning particle model from a variational problem. To do so, we must work with a classical mechanics system endowed with position variables $x^i$ and additional degrees of freedom. The latter are necessary to form the inner space associated with the particle, which, in turn, is described by a point in a world line \cite{Alexei_Bruno_Genaro_2012}. Once again, the reparametrization invariance plays a central role, as it will become clear along the discussion. The particular model presented here can be found in Refs. \cite{Deriglazov_2010, Deriglazov_2014}.   

Consider a phase space equipped with canonical Poisson bracket, for instance, $\{\omega^i,\pi_j\} = \delta^i{}_j$. Spin operators $\hat{S}_i$ act on the two-component wave function as $2\times2-$matrices, according to the definition
\begin{equation}
    \hat{S}_i := \frac{\hbar}{2}\sigma_i.
\end{equation}
This shows that the spin operators are proportional to the Pauli matrices
\[
\sigma_1 = \begin{bmatrix}
    0 & 1 \\
    1 & 0
\end{bmatrix}
,\;\;\;
\sigma_2 = \begin{bmatrix}
    0 & -i \\
    i & 0
\end{bmatrix}
,\;\;\;
\sigma_3 = \begin{bmatrix}
    1 & 0 \\
    0 & -1
\end{bmatrix}.
\]

Interesting properties follow through this definition. Most importantly,
\begin{equation}
    [\hat{S}_i,\hat{S}_j] = i\hbar\varepsilon_{ijk}\hat{S}^k
    \label{opcom}
\end{equation}
and
\begin{equation}
    \hat{S}\cdot\hat{S} = \hbar^2s(s+1) = \frac{3\hbar^2}{4}.
\end{equation}
In addition to that, spin operators form $SO(3)-$algebra with respect to the commutators (\ref{opcom}). Noting that this is different from the variables' algebra and that we have distinct numbers of variables and spin operators, we follow the common procedure of defining the spin as a vector product between coordinates of a phase space, say $\omega^i, \pi_j$. Thus,
\begin{equation}
    S_i := \varepsilon_{ijk}\omega^j\pi^k.
    \label{spindef}
\end{equation}
Poisson brackets for $S_i$ yield
\begin{align}
    \{S_i,S_j\} = \varepsilon_{ijk}S^k.
\end{align}
Then, the definition (\ref{spindef}) implies $SO(3)$-algebra for the spin variables. Besides that, we impose constraints to our variational problem,
\begin{align}
    \pi^i\pi_i = b^2,\;\;\;\omega^i\omega_i = a^2,\;\;\; \pi^i\omega_i = 0.
    \label{constraints}
\end{align}
Here, $a$ and $b$ are constants. This way, the number of degrees of freedom for the spinning particle coincides with the one chosen here in \eqref{constraints}, see \cite{Deriglazov_2010}.  

Our next step is to write a Lagrangian which gives, apart from dynamical equations, the imposed constraints (\ref{constraints}). Consider 
\begin{equation}
L_{spin} = \frac{1}{2g}(\dot{\omega}_i)^2 + \frac{1}{2}gb^2 - \frac{1}{\phi}(({\omega}_i)^2 - a^2).    
\end{equation}
$g(\tau)$ and $\phi(\tau)$ are auxiliary degrees of freedom. 
Firstly, we observe that the two initial terms above resemble those in \eqref{L.Free_Relativistic_Particle}. So, one expects a local symmetry due to the presence of first class constraints. In fact, the model is invariant under the following gauge transformations
\begin{align}
    \delta \omega_i = \xi \dot{\omega}_i,\;\;
    \delta g = (\xi g)^.,\;\;
    \delta \phi = \dot{\xi}\phi -\xi\dot{\phi},
    \label{symme}
\end{align}
where $\xi$ is an arbitrary function of $\tau$. The variables $\omega_i$ and $\pi_i$, which compose the spin, are affected by these local symmetries (generated by the constraints), while the quantities $S_i$ are left invariant under them. The case here is similar to what was discussed in the subsection \ref{kineticpov}. $S_i$ is gauge-invariant,  therefore, representing a possible observable quantity of the model.

Variation of $L_{spin}$ with respect to auxiliary variables gives
\begin{align}
    \frac{\delta L_{spin}}{\delta g} = 0 = b^2 - \frac{1}{g^2}(\dot{\omega}_i)^2 \Rightarrow g^2 = \frac{(\dot{\omega}_i)^2}{b^2}
    ,\label{var_g}\\
    \frac{\delta L_{spin}}{\delta\phi} = 0 = \frac{1}{\phi^2}((\omega_i)^2 - a^2) \Rightarrow (\omega_i)^2 = a^2.\label{var_phi}
\end{align}
From (\ref{var_g}) we see that $g(\tau)$ has no independent dynamics, while (\ref{var_phi}) implies that $\dot{\omega}^i\omega_i = 0$. Computing the conjugate momenta we get $\pi_i = \frac{\partial L_{spin}}{\partial\dot{\omega}^i} = \frac{\dot{\omega}^i}{g}$. Substitution of $g(\tau)$ into the last equality gives
\begin{equation}
    \pi_i = b\frac{\dot{\omega}^i}{\sqrt{(\dot{\omega}_i)^2}} \Rightarrow \pi_i^2 = b^2.
\end{equation}
Thus, we have obtained the desired constraints through variation of $L_{spin}$. This Lagrangian, however, can be presented in a more compact form, structurally close to the one in $S_1$, and still provide the same constraints. Since $g^2 = \frac{(\dot{\omega}_i)^2}{b^2}$, we can write
\begin{equation}
    L_{spin} = b\sqrt{(\dot{\omega}_i)^2} - \frac{1}{\phi}((\omega_i)^2 - a^2).
\end{equation}
We could go even further though. If we use the constraint $(\omega_i)^2 = a^2$ to exclude $\phi$ of the Lagrangian, and one of the $\omega$'s, say, $\omega_3$, the model would lose its manifest rotational invariance, since $\omega'_i = R_{ij}\omega_j$, $R_{ij} \in SO(3)$. However, it becomes fully reparametrization invariant, 
\begin{equation}
L_{spin} = b \sqrt{g_{\alpha\beta}(\omega_\gamma)\dot \omega^\alpha \dot \omega^\beta}; \,\, \alpha,\beta,\gamma =1,2.
\end{equation}
Here, $g_{\alpha\beta} = \delta_{\alpha\beta}+\frac{\omega_\alpha \omega_\beta}{a^2 - \delta_{\alpha\beta} \omega^a\omega^b}$ are the metric components over a sphere \cite{Abreu_Rizzuti_Mendes_Freitas_Nikoofard_2015} and this form of Lagrangian has the same very structure as the one presented in $S_1$.

At this point, we would like to highlight one of the main properties of the model above. It admits interaction with electromagnetic fields. Incorporating both position space variables and the spinning ones, together with the corresponding interaction, we find
\begin{multline}
   S = \int dt\Big(\frac{m}{2}\dot{x}^2_i + eA_i\dot{x}^i - eA_0 + \frac{1}{2g}(\dot{\omega}_i - \frac{e}{m}\epsilon_{ijk}v^jB^k)^2 
   + g\frac{b^2}{2a^2} - \frac{1}{\phi}(\omega^2_i - a^2)\Big).
    \label{bigaction}
\end{multline}
A computation shows that the canonical quantization leads to the Pauli equation \cite{Deriglazov_2014}. 

Below, we discuss a more complex spinning particle model, in which the invariance can also be accomplished.

\subsection{Staruszkiewicz model}

Let us now discuss another classical model that describes a relativistic rotator \cite{staruszkiewicz}. It is one of a series of papers concerning the description of spinning particles before and after canonical quantization, see \cite{kuzenko, BerezinMarinov, kassandrov2009, shirzad, alexei2014, Gamboa_Plyushchay_1998, Barut_1993, abgp} and references therein.

For this particular case, once again we have a reparametrization invariant formulation, with action given by
\begin{equation}\label{sm}
    S = \int d\tau m \sqrt{\dot{x}\eta\dot{x}}\sqrt{1 + \sqrt{-l^2\frac{\dot{k}\eta\dot{k}}{(k\eta\dot{x})^2}}}.
\end{equation}
The $x$-variables have usual meaning as a vector formed by pairs of events on Minkowski space, while $k$ is a null direction. According to the Wigner's idea that quantum systems can be classified according to irreducible representations of the Poincar\'e group, the parameters $m$ and $l$ were introduced by A. Staruszkiewicz and label the representations. In fact, defining the conjugate momenta $p_{\mu}= \partial_{\dot{x}^{\mu}}L$ and $\pi_{\mu}= \partial_{\dot{k}^{\mu}}L$, we can verify that the following Casimir invariants  are satisfied
\begin{eqnarray}\label{Casimir}
 p^{\mu}p_{\mu} = m^2,  \quad  \quad W^{\mu}W_{\mu} = - \frac{1}{4}m^4 l^{2}. 
\end{eqnarray}
Here, $W^{\mu}$ is the Pauli-Lubański pseudovector
\begin{eqnarray}
 W^{\mu} = - \frac{1}{2} \epsilon^{\mu \nu \alpha \beta}M_{\nu \alpha} p_{\beta}
\end{eqnarray}
and $M_{\nu \alpha}$ are the components of the total angular momentum, given by
\begin{eqnarray}
M_{\nu \alpha} = x_{\nu}p_{\alpha}-x_{\alpha}p_{\nu} + k_{\nu}\pi_{\alpha}-k_{\alpha}\pi_{\nu}.
\end{eqnarray}

The action (\ref{sm}) has a structure similar to that of $S_{1}$ and when we take $l=0$, the expression for free relativistic particle (\ref{sc}) is recovered. 

This is one among many examples where reparametrization invariance leads to pseudo-classical mechanics, with constraints in the corresponding Hamiltonian formulation. This particular model was used here because of its property of having Casimir invariants as parameters, instead of just constants of motion. A complete analysis of the model may be seen in \cite{kassandrov2009}. 

\section{Forceless Mechanics of Hertz}
\label{sec5}

Here we turn our attention to the opposite direction we have been following. We shall discuss one classical example where $S_2$ is explicitly used, while invariance under reparametrization is not present, the forceless mechanics of Hertz. It is a very interesting proposal because it can be seen as the first example where, through the usage of extra degrees of freedom, the force concept was substituted by the curved space itself. One observes a similar methodology in general relativity, proposed by A. Einstein.
Curiously enough, the model is of $S_2$-type but we can kill the extra degree of freedom though. We use geometrical arguments and physical reality instead of invoking local symmetries to characterize the observable sector of the theory. 

\subsection{Introducing the model}

Heinrich Hertz, who is greatly known for his experimental expertise, worked on a non-Newtonian form of mechanics. Philosophically, it seems that Hertz did not find satisfactory the idea that two distinct quantities, ``mass" and ``force", were necessary to explain the motion of a massive object \cite{hertz1899mechanics}. His idea, thus, was to describe the motion of any mechanical system as the motion of a free particle in a curved space. The curvature would appear in the model as a consequence of inserting auxiliary degrees of freedom to the system, which, in turn, would alter the space's metric. The idea here is close related to what Einstein proposed latter, in his general relativity theory. Gravitational interactions happen between mass objects in space-time due to the curve structure of the latter. This way, the concept of force is expendable.

In the Sections above, we have discussed some models whose dynamical equations are given by geodesics. However, the models had invariance under reparametrization and, because of that, the functions $x^i(\tau)$ had no physical meaning. Here, the goal is to present a model which resembles the action $S_2$, being the forceless mechanics of Hertz the chosen one. The system will have a well defined evolution parameter and, consequently, the dynamical equations shall describe a proper physical solution, except for the additional spacial dimension introduced by Hertz. We provide a modern detailed geometrical analysis of the model, in which the removal of the extra degree of freedom is naturally obtained. Our main result in this Section is the geometrical proof that guarantees the removal of the auxiliary degree of freedom.  

We begin by considering the standard Newtonian mechanics description of a particle in riemannian $m$-dimensional manifold $(\mathcal{M}^m, \delta)$, represented by $x^i(\tau)$, subject to a potential $\mathcal{U}(x^i)$. One can promptly write 
\begin{equation}
    S = \int d\tau\Big[\frac{1}{2}\delta_{ij} \dot{x}^i \dot{x}^j - \mathcal{U}(x^i)\Big]
    \label{S_potential}
\end{equation}
and, through variation of (\ref{S_potential}), obtain the equations of motion
\begin{equation}
    \ddot{x}^i = -\partial_i\mathcal{U}.
\end{equation}
Here, the indices $i,j,k$ go through $1$ to $m$.

To avoid the Newtonian force approach, the procedure developed by Hertz leads us to insert one extra function $X(\tau)$ in the set of functions representing the particle. Then, we begin to work with a $m+1$-dimensional differentiable manifold $\mathcal{M}^{m+1}$. Given a chart $(\varphi,U)$ and considering the indices $a,b = 1,...,m,m+1$, $x^a(\tau)$ is defined as
\begin{equation}
    (\varphi\circ\alpha):= x^a(\tau),
\end{equation}
where, as introduced before, $\alpha$ is the curve corresponding to the particle in the $\mathcal{M}^{m+1}$ manifold
\begin{align*}
    \alpha \,:\, I \subset \mathbb{R} &\rightarrow \mathcal{M}^{m+1}\\
     \tau &\mapsto \alpha(\tau).
\end{align*}

By using the (pseudo-)metric $g = \delta_{ij} d x^{i}\otimes d x^{j} + \frac{1}{2 \mathcal{U}(x^{i})} d X \otimes d X$, we consider the action functional 
\begin{align}
    S = \int d\tau \frac{1}{2}g_{ab}\dot{x}^a\dot{x}^b = \int d\tau\Big[\frac{1}{2}(\dot{x}^i)^2 + \frac{1}{4\mathcal{U}}\dot{X}^2\Big].\label{S_rewriten}
\end{align}
Hence, we obtained with (\ref{S_rewriten}) the action of a \textit{free particle} in the $\mathcal{M}^{m+1}$ manifold. We point out that the particle that \eqref{S_rewriten} describes is now immersed in a curved space, as the metric is explicitly $x^i$-dependent. Equations of motion for this particle assume the well-known geodesic form $\ddot x^a + \Gamma^a{}_{bc}\dot{x}^b \dot{x}^c=0$, reading separately 
\begin{equation}
    \ddot{x}^i + \frac{1}{4\mathcal{U}^2}\partial_i\mathcal{U}\dot{X}\dot{X} = 0
    \label{motion_1}
\end{equation}
and
\begin{equation}
    \ddot{X} - \frac{1}{\mathcal{U}}\partial_i\mathcal{U}\dot{x}^i\dot{X} = 0.
    \label{motion_2}
\end{equation}
The above equations allow us to relate the potential $\mathcal{U}$ with the connection coefficients $\Gamma(g)$, such as $\Gamma^{\,i}_{\,\,\,\,\,m+1,m+1} = \frac{1}{4\mathcal{U}^2}\partial_i\mathcal{U}$ and $\Gamma^{\,m+1}_{\,\,\,\,\,\,\,\,\,i,m+1} = -\frac{1}{\mathcal{U}}\partial_i\mathcal{U}$. Thus, by extending our configuration space, we were able to hide the potential $\mathcal{U}$ into $\Gamma(g)$, allowing one to interpret the classical potential as the cause of the curvature.

\subsection{Physical sector of the model}

Despite the fact this formalism allows us to work with a free particle, in a possibly more onerous space structure, we would still like to obtain the physical description of the system. In other words, while $x^a(\tau)$ describes a free particle in a curved space equipped with the metric $g$, the physical sector is still described by the functions $x^i(\tau)$, in the $\mathcal{M}^m$ manifold, with the Euclidean metric $\delta$. Let us now present some arguments to kill the auxiliary variable $X$, after constructing an appropriate model, to obtain the equivalent physical system.

First, we consider the classical approach of imposing initial conditions. Here, our generalized coordinates $x^i$ and $X$ obey
\begin{align}
    &x^i(0) = x^i_0,\;\;\;\;\;\;\dot{x}^i(0) = v^i_0,\\
    &X(0) = X_0,\;\;\;\;\;\;\dot{X}(0) = 2\mathcal{U}(x^i_0).
\end{align}

Equation (\ref{motion_2}) can be written as
\begin{equation}
    \frac{d}{d\tau}\Big[\frac{\dot{X}}{2\mathcal{U}}\Big]=0 \Rightarrow \dot{X} = 2k\mathcal{U}.
\end{equation}
Where $k$ is a constant. To respect the initial condition of $\dot{X}$, we must have $k = 1$. Substitution of $\dot{X} = 2\mathcal{U}$ into (\ref{motion_1}) gives
\begin{equation}
    \ddot{x}^i + \frac{1}{4\mathcal{U}^2}\partial_i\mathcal{U}\dot{X}\dot{X} = 0 \Rightarrow \ddot{x}^i + \partial_i\mathcal{U} = 0.
\end{equation}
That is the result of the initial formulation, independent of $X$. Here, we followed the tradition in Newtonian mechanics, which is to insert manually initial data, after we found how the system behaves dynamically. 

Another way to attack this problem is to set $k = \frac{\dot{X(0)}}{2\mathcal{U}(x^i(0))}$. Instead of fixing conditions that restrict $k = 1$, we write this constant in a way that it is now essentially part of the dynamical equations. The definition $\mathcal{U'} := k\mathcal{{U}}$ gives
\begin{equation}
    \ddot{x}^i = -\partial_i\mathcal{U'}.
\end{equation}
The parameter $k$ can be seen as a controller of the potential's strength.

At this point, we would like to emphasize that the formalism developed by Hertz has a non-uniqueness \cite{Wheeler}. In effect, consider
\begin{equation}
    L = \frac{1}{2}(\dot{x}^i)^2 + \frac{1}{4\mathcal{U}_1}\dot{X}^2_1 + \frac{1}{4\mathcal{U}_2}\dot{X}^2_2.
    \label{non_uniq_1}
\end{equation}
Equations of motion give
\begin{align}
    \ddot{x}^i &= -k^2_1\partial_i\mathcal{U}_1 - k^2_2\partial_i\mathcal{U}_2\\
               &= -k^2_1\partial_i(\mathcal{U}_1 + \gamma\mathcal{U}_2),
\end{align}
where $K_l = \frac{\dot{X_l(0)}}{2\mathcal{U}_l(x^i(0))}$, $l = 1,2$ and $\gamma = \Big(\frac{k_2}{k_1}\Big)^2$.

Alternatively, one can write
\begin{equation}
    L = \frac{1}{2}(\dot{x}^i)^2 + \frac{1}{\mathcal{U}_1 + \gamma\mathcal{U}_2}\dot{X}
\end{equation}
and obtain the same dynamical description for $x^i$, as we got from (\ref{non_uniq_1}). Thus, we may have different mechanical systems on the extended space described by the same equations of motion. These calculations indicate that, indeed, $X$ may not represent a physical variable.

As we have pointed out, when we extend our space, the particle behaves as if it were on a curved surface. Regarding $X(\tau)$ and its initial condition $X_0$, we see that equations for both $X$ and $x^i$ have been separated. As a consequence, the system's dynamics is not altered by different choices of $X_0$. Our next goal is to give geometrical proof to this claim.

\subsection{Killing Vector Fields}

We may regain the particle's physical trajectory by projecting the geodesic line described in the extended space on the $m-$dimensional space. Geometrically, we can prove this by showing that $\partial_X$ is one of the metric's Killing vector fields, that is, the vector fields which represent the direction of the symmetry of a manifold. In addition to that, we can show that $\partial_X$, under rotations, represents a privileged direction, which is not desirable if the isotropy of the space is taken for granted.

We begin our discussion with the very equation
\begin{equation}
    \mathcal{L}_Kg = 0,
\end{equation}
that represents the Lie Derivative of the metric tensor with respect to the Killing vector field $K$. It tells us that the metrical properties of the manifold $\mathcal{M}^{m+1}$ do not get altered by this vector field. Indeed, Denoting $g = g_{ab}(x^i)dx^a\otimes dx^b$, where, as before, $i,j,k \in \{1,\dots,m\}$ and $a,b,c \in \{1,\dots,m,m+1\}$, we have
\begin{align}
    \mathcal{L}_Kg &= \mathcal{L}_{K^c\partial_c}\big(g_{ab}dx^a\otimes dx^b\big) =\nonumber\\ &= K^c\partial_cg_{ab}dx^a\otimes dx^b + g_{ab}\partial_cK^adx^c\otimes dx^b + g_{ab}\partial_cK^bdx^a\otimes dx^c = \nonumber\\
    &= \Big(K^c\partial_cg_{ab} + g_{cb}\partial_aK^c + g_{ac}\partial_bK^c\Big)dx^a\otimes dx^b = 0.
\end{align}
Since $g_{m+1,m+1} = \frac{1}{2\mathcal{U}}$, the Killing equations are
\begin{align}
    &\frac{-K^i}{2\mathcal{U}}\partial_i\mathcal{U} + \partial_XK^X = 0,\label{killing_1}\\
    &\frac{1}{2\mathcal{U}}\partial_iK^X + \partial_XK^i = 0,\label{killing_2}\\
    &\partial_iK^j + \partial_jK^i = 0\label{killing_3}.
\end{align}
Substituting $K_1 = 0.\partial_i + 1.\partial_X$ in to the equations (\ref{killing_1},\ref{killing_2},\ref{killing_3}) one verifies that they are satisfied. In other words, $K_1$ is a Killing vector field of the metric $g$ which generates translations along the $\partial_X$ direction. 

Our next analysis consists of taking the vector field
\begin{equation}
    R^i_{m+1} = -x^i\partial_X + X\partial_i
    \label{Def.Rot.Fail}
\end{equation}
and inserting it in the Killing equations. It is clear that Eq. \eqref{killing_1} is not satisfied, since, generally, $\partial_i\mathcal{U} \neq 0$. This reveals that the rotation around fixed axes, including the subspace generated by $\partial_X$, breaks the space's isotropy. Curiously enough, this shows that a resultant force different than zero in the physical system implies $R^i_{m+1}$ is not a Killing vector field in the $\mathcal{M}^{m+1}$ manifold. However, there, the system is represented by a free particle, where the potential is hidden in the kinetic term. Thus, we could argue that $\partial_i\mathcal{U}$ is intrinsically affecting the extended system and it is responsible for the symmetry break in $R^i_{m+1}$'s direction. Meanwhile, it is easy to see that rotations given by $K_2 = -x^i\partial_j + x^j\partial_i$ obey the Killing equations. Regardless of having a resultant force on the system, these rotations do not affect the isotropic properties of the space. These results should suffice in showing that the direction of $\partial_X$ is privileged, advocating towards our previous claim that the auxiliary variable $X$ should not be part of our physical dynamical equations. Therefore, we have a geometrical proof that different trajectories on the extended space project on the same physical trajectory.

\section{Conclusions}
\label{con}

In this work we have made explicit the fundamental difference between two actions describing particles lying on a manifold.  While, in the first case, one minimizes an arc length, in the second we extreme a kinetic energy term. In the initial action, due to reparametrization invariance, the evolution parameter has no physical interpretation. This implies that not all degrees of freedom are the physical ones. It is reflected by the presence of a local symmetry generated by what is called a first class constraint in Dirac's terminology. On the other hand, even though in the second action the equations of motion read the same as in the previous one, the time evolution parameter does have physical interpretation, as it cannot be changed arbitrarily. There is no constraint into the formulation and all coordinates have physical interpretation in the sense that their dynamics is independent from one another. The two descriptions, however, turn out to be equivalent as we fix the condition $\dot x g \dot x=$const. 

We exposed the subtle difference of both actions with distinct examples. For the first case, we explored the action of a free relativistic particle with no internal structure and its natural continuation for a particle with spin. In all actions, invariance under reparametrizations plays a central role as fully global Lorentz invariance is explicit. It is worth mentioning that one may describe the free particle, with or without spin, by using physical coordinates. However the price you pay is to represent the Lorentz group non linearly.   

Our last Section was designed to present and analyze the forceless mechanics of Hertz. Following Hertz's procedure, we wrote the action of a free particle in curved space. Although force becomes an unrequired quantity to describe the system, as Hertz desired, we are still required to map the solution on the extended space to the physical one. Following this idea, we gave a few arguments to kill the auxiliary degree of freedom. Initially, it was shown that the extra variable vanishes with proper choices of initial conditions. It was also shown that proper adjustments of the parameter controlling the potential led to an analogous result. After this, the non-uniqueness of the Hertz formalism was highlighted.

One of the paper's main results was then presented. Using differential geometry, an elegant proof was given to justify the removal of the auxiliary degree of freedom. Analyzing the vector fields which broke the space's isotropy, we were able to determine that $\partial_X$, the vector field associated with the auxiliary variable, generates translations along its own direction. Thus, different choices for $X$ would map into the same physical solution, giving substantial basis for the projection argument. Additionally, it was exposed that rotations around fixed axes, which include the direction of $\partial_X$, are not Killing vector fields. As a consequence, a curious fact was pointed out. The presence of force in the physical sector implies a symmetry break on the extended space, even though we describe our system as a free particle on it. Particularly, it reveals that the extra degree of freedom should not take part on the physical dynamical solution of the system.

\section*{Acknowledgement}

This work is supported by Programa Institucional de Bolsas de Iniciação Científica - XXVII PIBIC/CNPq/UFJF-2018/2019, project number ID45246. M. A. Resende
thanks CAPES for the financial support.


\begin{thebibliography}{0}
	\bibitem{sen} R. Sen \textit{Causality, Measurement Theory and the Differentiable Structure of Space-Time} (Cambridge University Press, Cambridge, 2010)
	
	\bibitem{o2006elementary} B. O'Neill, Semi-Riemannian Geometry With Applications to Relativity (Academic Press, San Diego, 1983). 
	
	\bibitem{nakahara} M. Nakahara, Geometry, Topology and Physics (Institute of Physics Publishing, Bristol, 2003).
	
	\bibitem{alexeibook} A. A. Deriglazov, Classical Mechanics: Hamiltonian and Lagrangian Formalism (Springer International Publishing, Switzerland, 2017). 
	
	\bibitem{alexei}  A. A. Deriglazov, Potential motion in a geometric setting:
	presenting differential geometry methods in a classical mechanics course, Eur. J. Phys.
	29 (2008) 767.
	
	\bibitem{alexeipla}  A. A. Deriglazov, Classical-mechanical models without observable trajectories and the Dirac electron, Phys. Lett. A 377 (2012) 13-17. 
	
	\bibitem{bep} B. F. Rizzuti, E. M. C. Abreu, P. V. Alves, Electron structure through a classical description of the Zitterbewegung, Phys. Rev. D  90 (2014) 027502. 
	
	\bibitem{abgp} A. A. Deriglazov, B. F. Rizzuti, G. P. Zamudio and P. S. Castro, Non-Grassmann mechanical model of the Dirac equation,  J. Math. Phys.  53 (2012) 122303. 
	
	\bibitem{aap} A. A. Deriglazov and A. M. Pupasov-Maksimov, Geometric constructions underlying relativistic 
	description of spin on the base of non-Grassmann vector-like variable,  SIGMA 10 (2014) 012.
	
	\bibitem{kuzenko} S. M. Kuzenko, S. L. Lyakhovich and A. Yu Segal, A geometric model of the arbitrary spin massive particle,  Int. J. Mod. Phys. A 10 (1995) 1529-1552.
	
	\bibitem{staruszkiewicz} A. Staruszkiewicz, Fundamental relativistic rotator, Acta Phys. Polon. B  39 (2008) 109.
	
	\bibitem{barut1989geometry} A. O. Barut, Geometry and Physics: Non-Newtonian Forms of Dynamics (Bibliopolis, Napoli, 1989).
	
	\bibitem{Lusanna_2018} L. Lusanna, Dirac-Bergmann   constraints in physics: Singular Lagrangians, Hamiltonian constraints and the second Noether theorem, Int.  J. G.  Methods Mod. Phys.  15 (2018) 1830004.
	
	\bibitem{aadbfr1} A. A. Deriglazov and B. F. Rizzuti, Generalization of the extended Lagrangian formalism on a field theory and applications, Phys. Rev. D 83 (2011) 125011.
	
	\bibitem{Aharonov_Bohm_1959} Y. Aharonov and D. Bohm, Significance of electromagnetic potentials in the quantum theory, Phys. Rev. 115 (1959) 485.
	
	\bibitem{wthirring} W. Thirring, \textit{Classical Mathematical Physics - Dynamical Systems and Field Theories}(Springer-Verlag, New York, 1997). 
	
	\bibitem{Fulop_1999} G. G. Fulop, D. M. Gitman and I. V. Tyutin, Reparametrization invariance as gauge  symmetry, 
	Int. J. Theor. Phys. 38 (1999) 1941-1968. 
	
	\bibitem{Alexei_Bruno_Genaro_2012}  A. Deriglazov, B. Rizzuti and G. Zamudio, Spinning particles: possibility of space-time interpretation for the inner space of spin (Lap Lambert Academic Publishing, Saarbr\"ucken, 2012).
	
	\bibitem{Deriglazov_2010}  A. A. Deriglazov, From noncommutative sphere to nonrelativistic spin, 
	SIGMA 6 (2010) 016.
	
	\bibitem{Deriglazov_2014}  A. A. Deriglazov, Geometric constructions underlying relativistic description of spin on the base of non-grassmann vector-like variable, SIGMA 10 (2014) 012. 
	
	\bibitem{Abreu_Rizzuti_Mendes_Freitas_Nikoofard_2015} E. Abreu, B. Rizzuti, A. Mendes, M. Freitas and V. Nikoofard, Noncommutative and dynamical analysis in a curved phase-space, Acta Phys. Polon. 46 (2015) 879-904. 
	
	\bibitem{BerezinMarinov} F. A. Berezin and M. S. Marinov, Classical spin and Grassmann algebra,
	J. Eur. Theor. Phys.  Lett. 21 (1975) 320-321; Ann. Phys. 104 (1977) 336.
	
	\bibitem{Barut_1993} A. O. Barut and M. G. Cruz, Classical relativistic spinning particle with anomalous 
	magnetic moment: the precession of spin, J. Phys. A: Math. and Gener. 26 (1993) 6499-6506. 
	
	\bibitem{Gamboa_Plyushchay_1998} J. Gamboa and M. Plyushchay, Classical anomalies for spinning particles, 
	Nucl. Phys. B 512 (1998) 278-282.
	
	\bibitem{kassandrov2009} V. Kassandrov, N. Markova, G. Schaefer and A. Wipf, On a model of a classical relativistic particle of constant and universal mass and spin, J. Phys. A: Math. and Theor. 42 (2009) 315204. 
	
	\bibitem{alexei2014} A. A. Deriglazov, Lagrangian for the Frenkel electron, Phys. Lett. B 736 (2014) 278-282.
	
	\bibitem{shirzad} M. Hajihashemi and A. Shirzad, A generalized model for the classical relativistic spinning particle, Int. J.  Mod. Phys. A 31 (2016) 1650027. 
	
	\bibitem{hertz1899mechanics}  H. Hertz, The Principles of Classical Mechanics Presented in a New Form (Macmillan and Co., London, 1899). 
	
	\bibitem{Wheeler} W. Nicholas, Geometrical mechanics remarks commemorative of Heinrich Hertz, preprint.
	
	
\end{thebibliography}
\end{document}